\documentclass[twocolumn]{revtex4-1}

\usepackage{amsmath,hyperref,bbold,blkarray,mathtools,lipsum,placeins}

\begin{document}
	
	\title{Anomaly-free $U(1)^m$ extensions of the Standard Model}
	
	\author{Davi B. Costa}   
	
	\affiliation{University of S\~ao Paulo, S\~ao Paulo 05508-090, Brasil} 
	
	\begin{abstract}
		We construct anomaly-free $U(1)_1\times U(1)_2\times...\times U(1)_m$ gauge extensions of the Standard Model. To perform this construction we put together anomaly-free $U(1)$ extensions of one and two families of fermions. The availability of free parameters that enter linearly in the equations for the fermion charges and the large number of different classes of extensions may help other model builders interested in their use to solve problems of particle physics.
	\end{abstract}

	\maketitle
	
	\subsection{Introduction}
	
	Particle physics have problems that can be solved using gauged $U(1)^m\equiv U(1)_1\times U(1)_2\times...\times U(1)_m$ extensions of the Standard Model (SM) featuring multiple $Z^\prime$ bosons \cite{Langacker:2008yv}. 
	For $m=1$, some examples are dark matter \cite{Okada:2020evk,Okada:2016tci,Okada:2018tgy,Jho:2020sku,Choi:2020nan,Foot:2014uba,Dutta:2019fxn}, neutrino masses \cite{Asai:2018ocx,Bertuzzo:2018ftf,Nomura:2020twp,Das:2017deo,Das:2018tbd,Das:2019fee,Das:2018tbd,Das:2019pua} and the anomalous magnetic moment of the muon \cite{Heeck:2011wj,Allanach:2015gkd,Amaral:2020tga}. 
	However,  in order to be well behaved at high energies a gauge symmetry must be free of anomalies \cite{Adler:1969gk,Bardeen:1969md,Gross:1972pv,Georgi:1972bb}. General solutions to the anomaly cancellation equations are not always useful because
	free parameters may be trapped in complicated polynomial expressions. However, in composite gauge theories, there is a subset of fermions that independently cancel anomalies, and one can use these different subsets to add free parameters that are easily available because they enter linearly in the equations for fermion charges.
	
	In this paper we will construct composite anomaly-free $U(1)^m$ gauge extensions of the SM. In sections \ref{B} and \ref{C} we will settle notation and define concepts, in \ref{D} and \ref{E} we will make the construction using one-family and two-families anomaly-free $U(1)$ extensions as building blocks in each case, and in \ref{G} we will present other extensions that can be made with additional charged Weyl fermions.
	
	\subsection{Anomaly equations }\label{B}
	
	Let $[z_{ij}^\ell]$ denote the charges of the SM fermions under an $U(1)^m=U(1)_1\times U(1)_2\times...\times U(1)_m$ gauge symmetry. The index $i\in\{1,2,3\}$ refers to the three families, $\ell\in\{1,2,...,m\}$ to the $U(1)_\ell$ subgroup and $j\in\{1,2,3,4,5,6\}\mapsto\{N,Q,D,L,U,E\}$ to the fermion fields which are in the following representations of the SM gauge group $G_{\textrm{SM}}\equiv SU(3)\times SU(2)\times U(1)_Y$:
	
	\begingroup
\setlength{\tabcolsep}{10pt} 
\renewcommand{\arraystretch}{0.75} 
\begin{table}[ht]
	\centering
	\begin{tabular}{cccc}
		Fermion fields &  SU(3) & SU(2) & $U(1)_Y$ \\ \hline
		N & \textbf{1} & \textbf{1} & 0 \\ 
		Q & \textbf{3} & \textbf{2} & 1 \\ 
		D & $\overline{\textbf{3}}$ & \textbf{1} & 2\\
		L & \textbf{1} & \textbf{2} & -3\\
		U & $\overline{\textbf{3}}$ & \textbf{1} & -4\\
		E & \textbf{1} & \textbf{1} & 6\\ \hline
	\end{tabular}
	\label{fermionfields}
\end{table}
\FloatBarrier
\endgroup

    \noindent We assumed that right-handed neutrinos exist to generalize our results and to simplify their exposition. This assumption does no harm, as we can assign zero charges under the extended $U(1)'s$ to them so that effectively they do not exist. Following the conventions of previous work \cite{Costa:2019zzy,Costa:2020dph}, fermions are taken as left-handed fields which can be done via charge conjugation. We also assume that $U(1)$ charges are always given for a normalization of the gauge couplings where all charges are integers with no common divisor and the biggest charge, in absolute value, is positive. Often we will use the notation $[z_{ij}^\ell]$ with the $i$ or $\ell$ indices suppressed. If the $i$ index is suppressed we will be considering a single family of the SM, if the $\ell$ index is suppressed the $U(1)$ group being considered will be clear in the context. In this paper, $m$ is always a given natural number.
	
	With this notation and convention the hypercharges $[y_{ij}]$ of the SM fermions, referred from here on with the letter $y$, are  $y_{i1}=0$, $y_{i2}=1$, $y_{i3}=2$, $y_{i4}=-3$, $y_{i5}=-4$ and $y_{i6}=6$ for all $i\in\{1,2,3\}$. We can also present it as a 3-tuple of $[y_j]\equiv[0,1,2,-3,-4,6]$ or in matrix form
	
	\begin{align}
	[y_{ij}]=\begin{bmatrix} y_j \\ y_j \\ y_j \end{bmatrix}=\hspace{2mm}\begin{blockarray}{ccccccc}
	N & \hspace{1mm} Q & \hspace{1mm} D & \hspace{2mm} L & \hspace{2mm} U & \hspace{1mm} E  \\
	\begin{block}{[cccccc]c}
	0 & 1 & 2 & -3 & -4 & 6 & \hspace{2mm} 1st\\
	0 & 1 & 2 & -3 & -4 & 6 & \hspace{2mm} 2nd\\
	0 & 1 & 2 & -3 & -4 & 6& \hspace{2mm} 3rd\\ 
	\end{block}\\
	\end{blockarray}.
	\label{hypercharge}
	\end{align}
	
	Without adding new fermions charged under hypercharge, an $U(1)$ gauge extension of the SM, $G_{\textrm{SM}}\times U(1)$, is \textit{anomaly-free} if the charges $[z_{ij}]$ associated with the additional $U(1)$ subgroup satisfies the following system of diophantine equations:
	
	\begin{subequations}
	\begin{align}
	& 0=\sum_{i,j}n_jz_{ij}^3, \label{firstanomalyequation}\\
	& 0=\sum_{i,j}n_jz_{ij}y_{ij}^2\label{U(1)U(1)_Y^2}, \\
	& 0=\sum_{i,j}n_jz_{ij}^2y_{ij},\\
	& 0=\sum_{i,j}\delta_j^{grav}n_jz_{ij},\\
	& 0=\sum_{i,j}\delta_j^{su(2)}n_jz_{ij},\label{anomalySU(2)}\\
	& 0=\sum_{i,j}\delta_j^{su(3)}n_jz_{ij} \label{lastanomalyequation},
	\end{align}
	\end{subequations}
	
	\noindent where $n_j$ is the number of Weyl fermions in the $j$ multiplet, and $\delta_j^s$ is equal $1$ if multiplet $j$ is charged under $s$ and $0$ otherwise, with $s\in\{grav,su(2),su(3)\}$.  Using vector notation we have $[n_j]=[1,6,3,2,3,1]$, $[\delta_j^{grav}]=[1,1,1,1,1,1]$, $[\delta_j^{su(2)}]=[0,1,0,1,0,0]$ and $[\delta_j^{su(3)}]=[0,1,1,0,1,0]$. More generally, an $U(1)^m=U(1)_1\times U(1)_2\times...\times U(1)_m$ gauge extension of the SM, $G_{\textrm{SM}}\times U(1)^m$, is anomaly-free if the charges $[z_{ij}^\ell]$, $[z_{ij}^{\ell^{\prime}}]$ and $[z_{ij}^{\ell^{\prime\prime}}]$ associated with the subgroups $U(1)_\ell,U(1)_{\ell^\prime}$ and $U(1)_{\ell^{\prime\prime}}$ satisfies (\ref{firstanomalyequation})-(\ref{lastanomalyequation}) and
		
	\begin{align}
	0=\sum_{i,j}n_jz_{ij}^\ell z_{ij}^{\ell^\prime}z_{ij}^{\ell^{\prime\prime}}.
	\label{U(1)U(1)U(1)}
	\end{align}
	
	\noindent for all $\ell,\ell^\prime,\ell^{\prime\prime}\in\{1,2,...,m\}$. These definitions are sufficient because an $G_{\textrm{SM}}\times U(1)^m$ gauge theory with no $SU(2)$ anomaly \cite{Witten:1982fp,Wang:2018qoy} does not have other global anomalies \cite{Davighi:2019rcd,Wan:2019fxh,Wang:2018cai,Garcia-Etxebarria:2018ajm}. In this paper, we will be refering to \textit{one-family} and \textit{two-families} anomaly-free extensions, these are defined in the same way as a generic anomaly-free $U(1)^m$ extension but for a vector $[z_j]$ and a $2-$tuple of vectors.
	
	
	Recently, an atlas with solutions for the anomaly equations
 	(\ref{firstanomalyequation})-(\ref{lastanomalyequation}) was constructed \cite{Allanach:2018vjg}. Then a general solution was found for the first two equations \cite{Costa:2019zzy}, and for this solution, it was given a geometric interpretation \cite{Allanach:2019gwp}. While the present work was being written, the full system of equations was solved using this geometric method \cite{Allanach:2020zna}. Other aspects of this system of equations including (\ref{U(1)U(1)U(1)}) were also explored \cite{Costa:2020dph,Allanach:2019uuu,Batra:2005rh}. 
	
	Generalizing a notion presented in \cite{Costa:2020dph}, we say that a gauge theory is \textit{composite} if there is a non-empty proper subset of fermions whose charges independently satisfy the anomaly equations. In particular, the SM is composite because each family of fermions is independently anomaly-free. In fact, the only non-empty proper subsets of fermion fields of the SM whose hypercharge independently satisfy all anomaly equations are those that make a complete family. Therefore, without adding new Weyl fermions charged under hypercharge, all composite $U(1)^m$ extensions of the SM are made with $U(1)$ extensions that are one-family or two-families anomaly-free. This implication is easy to prove by contradiction. First, assume that there is an $U(1)^m$ extension that does not satisfy the condition, then there would be a non-empty proper subset of fermions of the SM such that anomaly equations cancel independently which is not one or two families, but there is none.
	
		

	\subsection{Family permutation}\label{C}
	
	Before presenting the one-family and two-families anomaly-free extensions that we will use as building blocks, we will formalize a notion that will help in the exposition of our results. Given a collection of charges $[z_{ij}]$ and $\sigma=(\sigma_1,\sigma_2,\sigma_3,\sigma_4,\sigma_5,\sigma_6)\in S_3^6$ where each $\sigma_j\in S_3$ is a permutation of $\{1,2,3\}$, we define the \textit{family permutation of $[z_{ij}]$ by $\sigma$} as the new collection
	
	\begin{align}
	\sigma[z_{ij}]\equiv[z_{ij}^\prime]=[z_{\sigma_j(i)j}].
	\label{familypermutation}
	\end{align}
	
	\noindent In words, the family permutation of $[z_{ij}]$ permutes the charges of each multiplet among the families in an independent way. With this definition, the family universality of the hypercharge can be stated precisely as
	
	\begin{align}
	    [y_{ij}]=\sigma[y_{ij}]
	    \label{familyuniversal}
	\end{align}
	
	\noindent for all $\sigma\in S_3^6$. This symmetry suggests redundancy in the way fermions are partitioned into three families.
	
	Concerning $U(1)$ extensions of the SM we have the following proposition: the family permutation of an anomaly-free $U(1)$ extension is also anomaly-free. To be precise, if $[z_{ij}]$ satisfy the anomaly equations (\ref{firstanomalyequation})-(\ref{lastanomalyequation}), then $\sigma[z_{ij}]$ also satisfy them for all $\sigma\in S_3^6$. With the notation introduced it is straightforward to verify this proposition. For example, if the charges $[z_{ij}]$ satisfy the anomaly equation (\ref{U(1)U(1)_Y^2}) then the charges $[z_{ij}^\prime]=\sigma[z_{ij}]$ satisfy the same equation since
	
    \begin{align}
    \sum_{i,j} g_jz_{ij}^{\prime}y_{ij}^2=\sum_{i,j}g_jz_{\sigma_j(i)j}y_{\sigma_j(i)j}^2=0,
    \end{align}
    
    \noindent where I used the fact that the hypercharge is family universal (\ref{familyuniversal}). More generally, if we have an anomaly-free $U(1)^m$ extension of the SM with charges $[z_{ij}^\ell]$, then the extension associated to $\sigma[z_{ij}^\ell]$ is also anomaly-free. 
	


	If all $U(1)$ subgroups of $U(1)^m$ are family universal, the proposition presented above is of no interest since $\sigma[z_{ij}^\ell]=[z_{ij}^\ell]$ for all $\sigma\in S_3^6$ and $\ell\in\{1,2,...,m\}$. However, if there is an $U(1)_k$ subgroup that is not family universal for some $1\leq k\leq m$, then in general $\sigma[z_{ij}^k]\neq[z_{ij}^k]$, and we can obtain different extensions by doing family permutations. To be exact, for each anomaly-free $U(1)^m$ extension there is a total of $|S_3^6|=6^6=46656$ other anomaly-free extensions. 
	
	

	\subsection{One-family anomaly-free extensions}\label{D}

	The one-family anomaly-free $U(1)$ extensions of the SM are
	
	\begin{equation}
	[f_j(\vec{z})]=[z_1, z_2, 2 z_2 - z_1, -3 z_2, z_1 - 4 z_2, 6 z_2 - z_1],
	\label{onefamily}
	\end{equation}	
	
	\noindent for all $\vec{z}=(z_1,z_2)\in\mathbb{Z}^2$. Recall our convention for the $j$ indices, namely $j\in\{1,2,3,4,5,6\}\mapsto\{N,Q,D,L,U,E\}$, and see \cite{Allanach:2018vjg} for details on how to derive this result. When $z_1=0$ the charges are proportional to one family of hypercharge, when $\vec{z}=(1,0)$ they are proportional to the charges associated to $U(1)_{T3R}$, and to $U(1)_{B-L}$ when $\vec{z}=(3,1)$.
	
	A surprising fact is that an $U(1)^m$ gauge theory with the same fermionic content as one family of the SM and with charges given by (\ref{onefamily}) with different parameters for each $U(1)$ component, is anomaly-free. Formally, the charges $[z_j^\ell]=[f_j(\vec{z}^{\hspace{0.3mm}\ell})],[z_j^{\ell^\prime}]=[f_j(\vec{z}^{\hspace{0.3mm}\ell^{\prime}})]$ and $[z_j^{\ell^{\prime\prime}}]=[f_j(\vec{z}^{\hspace{0.3mm}\ell^{\prime\prime}})]$ associated with the $U(1)_\ell$, $U(1)_{\ell^\prime}$ and $U(1)_{\ell^{\prime\prime}}$ subgroups satisfies 
	
	\begin{align}
	\sum_{j}n_jz_j^{\ell}z_j^{\ell^{\prime}}z_j^{\ell^{\prime\prime}}=\sum_jn_jf_j(\vec{z}^{\hspace{0.3mm}\ell})f_j(\vec{z}^{\hspace{0.3mm}\ell^{\prime}})f_j(\vec{z}^{\hspace{0.3mm}\ell^{\prime\prime}})=0,
	\label{onefamily3}
	\end{align}
	
	\noindent for any $\vec{z}^{\hspace{0.3mm}\ell},\vec{z}^{\hspace{0.3mm}\ell^\prime},\vec{z}^{\hspace{0.3mm}\ell^{\prime\prime}}\in\mathbb{Z}^2$ and all $\ell,\ell^\prime,\ell^{\prime\prime}\in\{1,2,...,m\}$. This does not seem to be trivial because (\ref{onefamily}) is found as a solution to (\ref{firstanomalyequation})-(\ref{lastanomalyequation}) and nothing else.
	
	A consequence of this fact is that an $U(1)^m$ extension of the SM with the charges of each subset of fermions that make a complete family under each $U(1)$ given by a multiple of (\ref{onefamily}) satisfies all anomaly equations. To be more precise, given $m\in\mathbb{N}$ and $\sigma\in S_3^6$, an $U(1)^m=U(1)_1\times U(1)_2\times...\times U(1)_m$ extension with the charges associated with the $U(1)_\ell$ subgroup given by
	
	\begin{align}
	[z_{ij}^\ell]\equiv\sigma[k_i^\ell f_j(\vec{z}_i^{\hspace{0.3mm} \ell})]=\sigma\left[\begin{array}{c}
	k_1^\ell f_j(\vec{z}_1^{\hspace{0.3mm} \ell})\\
	k_2^\ell f_j(\vec{z}_2^{\hspace{0.3mm} \ell})\\
	k_3^\ell f_j(\vec{z}_3^{\hspace{0.3mm} \ell})\\
	\end{array}\right],
	\label{onefamilyU(1)^m}
	\end{align}
	
	 \noindent with $\vec{z}_1^{\hspace{0.3mm}\ell},\vec{z}_2^{\hspace{0.3mm}\ell},\vec{z}_3^{\hspace{0.3mm}\ell}\in\mathbb{Z}^2$ and $k_1^\ell,k_2^\ell,k_3^\ell\in\mathbb{Z}$ for all $\ell\in\{1,2,...,m\}$, is anomaly-free. Note that $\sigma$ does not depend on $\ell$ because otherwise the anomaly equation (\ref{U(1)U(1)U(1)}) would not be satisfied in general. These extensions are composite because anomalies cancel for subsets of fermions that make a complete family, for $\sigma=id$, the subsets are the actual families. Finally, these extensions have free parameters that enter linearly in the formulas for the fermion charges as stated in the introduction. 
	
	\subsection{Two-families anomaly-free extensions}\label{E}
	
	The two-families anomaly-free $U(1)$ extensions of the SM are
	
	\begin{align}
	\left[\begin{array}{c}
		h_{1j}(\vec{z},\vec{x})\\
		h_{2j}(\vec{z},\vec{x})
	\end{array}\right]=\left[\begin{array}{c}
	f_j(\vec{z})+g_j(\vec{x})\\
	f_j(\vec{z})-g_j(\vec{x})
	\end{array}\right],
	\label{twofamilies}
	\end{align}
	
	\noindent with $[f_j(\vec{z})]$ defined as in (\ref{onefamily}) and $[g_j(\vec{x})]$ parameterized by 4 integer parameters depending on $\vec{z}$. If $z_1=0$, right-handed neutrinos will form a vector-like pair with arbitrary charge, and 

	\begin{subequations}
		\begin{align}
		&g_Q(\vec{x}) =-(x_1^2- x_2^2 - x_3^2 + 2 x_4^2 ),\\
		&g_D(\vec{x}) =2x_1x_2,\\
		&g_L(\vec{x}) =-(x_1^2+ x_2^2+ x_3^2 - 2 x_4^2),\\
		&g_U(\vec{x}) =2x_1x_4,\\
		&g_E(\vec{x})=2 x_1x_3,\hspace{51mm}		
		\end{align}
	\end{subequations}
	
	 \noindent for any $\vec{x}=(x_1,x_2,x_3,x_4)\in\mathbb{Z}^4$. If $z_1\neq0$ then
	
	\begin{subequations}
	\begin{align}
	& g_N(\vec{x}) = 2 x_1 (x_2^2 + 3 x_3^2 - 3 x_4^2),\\
	& g_Q(\vec{x}) = -x_1^2 + x_2^4 + 9 (x_3^2 -x_4^2)^2 - 2x_2^2 (x_3^2 + x_4^2),\\
	& g_D(\vec{x}) = 4 x_1 x_2 x_3,\\	
	& g_L(\vec{x}) = -x_1^2 - x_2^4 - 9 (x_3^2 - x_4^2)^2 + 2 x_2^2 (x_3^2 + x_4^2),\\
	& g_U(\vec{x}) = 4 x_1 x_2 x_4,\\
	& g_E(\vec{x}) = 2 x_1(x_2^2 - 3 x_3^2 + 3 x_4^2),
	\end{align}	
	\end{subequations}

	\noindent for any $\vec{x}=(x_1,x_2,x_3,x_4)\in\mathbb{Z}^4$.  Again, recall our convention for the $j$ indices, $j\in\{1,2,3,4,5,6\}\mapsto\{N,Q,D,L,U,E\}$, and see \cite{Allanach:2018vjg} for details on how to derive these results. The first family of solutions is proportional to the charges associated with $U(1)_{L\mu-L\tau}$ for $z_2=x_3=x_4=0$ and $x_1=-x_2$. 
	
	Unfortunately, an $U(1)^2$ gauge theory with the same fermionic content as two families of the SM and charges given by (\ref{twofamilies}) with different parameters for each $U(1)$ is not in general anomaly-free. This means that we cannot use them, without other restrictions, to build $U(1)^m$ extensions as we did in (\ref{onefamilyU(1)^m}). However, an $U(1)^3$ gauge theory with the charges associated with each $U(1)$ component given by
	
	\begin{align}
	\left[\begin{array}{c}
	h_{1j}(\vec{z_1},\vec{x})\\
	h_{2j}(\vec{z_1},\vec{x})
	\end{array}\right],\quad
	\left[\begin{array}{c}
	f_j(\vec{z_2})\\
	f_j(\vec{z_2})
	\end{array}\right],\quad\textrm{and}\quad\left[\begin{array}{c}
	f_j(\vec{z_3})\\
	f_j(\vec{z_3})
	\end{array}\right],
	\label{2‑tuples}
	\end{align}
	
	\noindent is anomaly-free for all $\vec{z}_1,\vec{z}_2,\vec{z}_3\in\mathbb{Z}^2$ and $\vec{x}\in\mathbb{Z}^4$, with the additional condition that if $z_{11}=0$ then $z_{21}=z_{31}=0$. The anomaly equations (\ref{U(1)U(1)U(1)}) that are linear with (\ref{twofamilies}) are satisfied because (\ref{onefamily3}) and $h_{1j}(\vec{z},\vec{x})+h_{2j}(\vec{z},\vec{x})=2f_j(\vec{z})$. But, it seems that it is not trivial that 
	
	\begin{align}
	\sum_{j}n_jf_j(\vec{z_2})(h_{1j}(\vec{z_1},\vec{x})^2+h_{2j}(\vec{z_1},\vec{x})^2)=0,	
	\end{align}

	\noindent for all $\vec{z_1},\vec{z_2}\in\mathbb{Z}^2$ and $\vec{x}\in\mathbb{Z}^4$, with the additional condition that if $z_{11}=0$ then $z_{21}=0$.
	
	Similarly to (\ref{onefamilyU(1)^m}), a consequence of this fact is that an $U(1)^m$ extension of the SM with the charges of one family under each $U(1)$ given by a multiple of (\ref{onefamily}) and the charges of the other two families under each $U(1)$ given by a multiple of one among (\ref{2‑tuples}), is anomaly-free. To be precise, given $m\in\mathbb{N}$, $\vec{z}\in\mathbb{Z}^2$, $\vec{x}\in\mathbb{Z}^4$ and $\sigma\in S_3^6$, an $U(1)_1\times U(1)_2\times...\times U(1)_m$ extension with charges $[z_{ij}^\ell]$  associated with the $U(1)_\ell$ subgroup given by
	
	\begin{align}
	\sigma\left[\begin{array}{c}
	k_1^\ell f_j(\vec{z}_1^{\hspace{0.5mm\ell}})\\
	k_2^\ell h_{1j}(\vec{z},\vec{x})\\
	k_2^\ell h_{2j}(\vec{z},\vec{x})
	\end{array}\right]\quad\textrm{or}\quad\sigma\left[\begin{array}{c}
	k_1^\ell f_j(\vec{z}_1^{\hspace{0.5mm\ell}})\\
	k_2^\ell f_j(\vec{z}_2^{\hspace{0.5mm\ell}})\\
	k_2^\ell f_j(\vec{z}_2^{\hspace{0.5mm\ell}})\\
	\end{array}\right]
	\label{U(1)^m}
	\end{align}
	
	\noindent with $k_1^\ell,k_2^\ell\in\mathbb{Z}$ and $\vec{z}_1^{\hspace{0.3mm}\ell},\vec{z}_2^{\hspace{0.3mm}\ell}\in\mathbb{Z}^2$ for all $\ell\in\{1,2,...,m\}$, is anomaly-free. Again, this statement is true if $z_{1}=0\Rightarrow z_{21}^\ell=0$ for all $\ell\in\{1,2,...,m\}$. Note that $\sigma,\vec{z},\vec{x}$ does not depend on $\ell$, because otherwise the cross anomaly equations (\ref{U(1)U(1)U(1)}) would not be satisfied in general.
	
	If we have $2\rightarrow3$ in the third line of the second collection in (\ref{U(1)^m}) and if $\vec{z},\vec{x}$ had a dependence on $\ell$, then we would have specified all potential composite $U(1)^m$ extensions of the SM. However, as we mentioned previously, the cross anomaly equations are not satisfied in general this way. It is possible to find relations between the parameters for the equations to be satisfied in particular cases, but it is hard to construct an $U(1)^m$ extension for generic $m\in\mathbb{N}$ from them. In conclusion, (\ref{onefamilyU(1)^m}) and (\ref{U(1)^m}) comprise a huge and particularly simple part of all composite anomaly-free $U(1)^m$ extensions of the SM.
	
	\subsection{Some composite extensions with new fermions}\label{G}
	
	So far we have presented $U(1)^m$ extensions of the SM without additional fermions. A simple way to add new fermions to this extensions without spoiling anomaly cancellation is by having them independently anomaly-free for $G_{\textrm{SM}}\times U(1)^m$. The minimal case is to add $2$ fermions which are singlets of $SU(3)\times SU(2)$ with opposite charges under each $U(1)$. In this case the resulting theory will be vector-like. In order to be chiral, it is necessary to add at least $5$ Weyl fermions \cite{Costa:2020dph}. In a sense, the simplest example is to have their $U(1)$ charges equal to $k^\ell[9, -8, -7, 5, 1]$ with $k^\ell\in\mathbb{Z}$ for all $\ell\in\{0,1,2,...,m\}$ where I used $\ell=0$ to refer to $U(1)_Y$. In these type of constructions the additional fermions are independently anomaly-free, however, there are other constructions in which this is not the case, and among them, there is one such that it is necessary to add only 4 Weyl fermions to have a chiral theory. 
	
	Any composite anomaly-free extension of the SM is made either of one-family (\ref{onefamily}) or two-families (\ref{twofamilies}) anomaly-free $U(1)$ extensions because the $i$ index is the only way to partition the sum of anomaly equations for hypercharge in pieces that are independently equal to zero. However, if we break the SM fields in their Weyl components there are other ways to do this. The other possibilities are:
    
    \pagebreak
	\begin{widetext}
	\begin{subequations}
	\begin{align}
	& \left[\begin{array}{cccccccccccccccccccc}
	6 & -4 & -4 & -4 & -4 & 2 & 2 & 2 & 2 & 2 &  \\ 
	6 &  -4 & -4 & -4 & -3 & -3 &  2 & 2 & 2 &  1 & 1 & 1 & 1 & 1 & 1 &   \\
	6 & -4 & -4 &  -3 & -3 & -3 & -3 & 2 & 1 & 1 & 1 & 1 & 1 & 1 & 1 & 1 & 1 & 1 & 1 & 1
	\end{array}\right],\label{best}\\
	& \left[\begin{array}{ccccccccccccccccccccccccccccccc}
	6 & -4 & -4 & -4 & -3 & -3 & -3 & 2 & 2 & 2 & 2 & 2 & 2 & 2 & 1 &\\
	6 & 6 & -4 & -4 & -4 & -4 & -4 & -4 & -3 & -3 & -3 & 2 & 2 & 1 & 1 & 1 & 1 & 1 & 1 & 1 & 1 & 1 & 1 & 1 & 1 & 1 & 1 & 1 & 1 & 1 \\
	\end{array}\right],
	\label{1530}\\
	& \left[\begin{array}{cccccccccccccccccccccccccccccc}
	6 & -4 & -4 & -3 & -3 & -3 & -3 & -3 & 2 & 2 & 2 & 2 & 2 & 1 & 1 & 1 & 1 & 1 & 1 & 1\\
	6 & 6 & -4 & -4 & -4 & -4 & -4 & -4 & -4 & -3 & 2 & 2 & 2 & 2 & 1 & 1 & 1 & 1 & 1 & 1 & 1 & 1 & 1 & 1 & 1\\
	\end{array}\right],\\
	& \left[\begin{array}{cccccccccccccccccccccccccccccc}
	6 & -4 & -4 & -3 & -3 & -3 & -3 & -3 & -3 & 2 & 2 & 2 & 2 & 2 & 2 & 2 & 2 & 2 & 1 & 1\\
	6 & 6 & -4 & -4 & -4 & -4 & -4 & -4 & -4 & 1 & 1 & 1 & 1 & 1 & 1 & 1 & 1 & 1 & 1 & 1 & 1 & 1 & 1 & 1 & 1\\
	\end{array}\right].
	\label{2025}
	\end{align}
	\end{subequations}
	\end{widetext}
	
	We regard these structured collections of charges as matrices with null entries and denote them by $[z_{kl}]$. The order in which the elements of each row are organized does not have any meaning. These structured collections of charges are all made with the hypercharges of the 45 charged Weyl fermions of the SM, but the k and l indices do not denote family and fermion field as before. These are partitions of the SM hypercharges such that for each $k$ we have	
	
	\begin{subequations}
	\begin{align}
	0 & = \sum_{l} z_{kl},
	\label{anomalyU(1)}
	\\ 
	0 & = \sum_{l} z_{kl}^3.
	\end{align}
	\end{subequations}

	These partitions can be used to construct different classes of composite anomaly-free extensions of the SM by finding $U(1)$ extensions to the different rows as we did in the previous sections for the family partition. But these extensions will not respect the full gauge structure of the SM because Weyl fermions are not built-in multiplets. However, we can solve this problem by adding pairs of Weyl fermions which are vector-like under $U(1)_Y$ although chiral under $U(1)_Y\times U(1)$. We will illustrate this construction for (\ref{best}) which is the partition that needs a smaller number of additional Weyl fermions. We will consider the simplest $U(1)$ extension, which is the repetition of a multiple of hypercharge in the new $U(1)$ which turns the cross anomaly equations into the cubic.
	
	First, we add to the Standard Model four Weyl fermions $\psi_1,\psi_2,\psi_3$ and $\psi_4$ which are singlets of $SU(2)\times SU(3)$ and have hypercharges respectively equal to $4,-4,2,-2$. Starting with $\{6,-4,-4,-4,-4,2,2,2,2,2\}$ of (\ref{best}), we first append the pair $\{2,-2\}$ to it. Now, we gather three $-4$'s in one $U$ multiplet and the six $2$'s in two $D$'s.  Then, we attribute $6$ to $E$, $-4$ to $\psi_2$ and $-2$ to $\psi_4$. For concreteness, the charges of the Standard Model fermions under this new $U(1)$ can be
	
	\begin{align}
	[v_{ij}]=\hspace{2mm}\begin{blockarray}{ccccccc}
	N & \hspace{1mm} Q & \hspace{1mm} D & \hspace{2mm} L & \hspace{2mm} U & \hspace{1mm} E  \\
	\begin{block}{[cccccc]c}
	0 & 0 & 2 & 0 & -4 & 6 & \hspace{2mm} 1st\\
	0 & 0 & 2 & 0 & 0 & 0 & \hspace{2mm} 2nd\\
	0 & 0 & 0 & 0 & 0 & 0& \hspace{2mm} 3rd\\ 
	\end{block}\\
	\end{blockarray},
	\label{uextension}
	\end{align}
	
	\noindent and the additional chiral set with four Weyl fermions can have $U(1)_Y\times U(1)$ charges equal to
	
	\begingroup
\setlength{\tabcolsep}{10pt} 
\renewcommand{\arraystretch}{1} 
    \begin{table}[ht]
	\begin{tabular}{ccc}
	\textrm{Fermions fields} & $U(1)_Y$ & $U(1)$ \\ \hline
	$\psi_1$ & 4 & 0 \\
	$\psi_2$ & -4 & -4\\
	$\psi_3$ & 2 & 0\\
	$\psi_4$ & -2 & -2\\ \hline
	\end{tabular}
	\end{table}
	\endgroup
	
	This composite extension satisfy all anomaly equations associated with $U(1)_Y\times U(1)$ by construction and somewhat surprisingly also satisfy the (\ref{anomalySU(2)}) and (\ref{lastanomalyequation}) anomaly cancellation equations associated with $SU(3)$ and $SU(2)$
	
    \begin{align}
    & \sum_{i,j}g_j\delta^{su(2)}_ju_{ij}=0,\\
    & \sum_{i,j}g_j\delta^{su(3)}_ju_{ij}=3\times(-4)+2\times 3\times 2=0.
    \end{align}
	
	\noindent Note that the additional Weyl fermions are vector-like under $U(1)_Y$ but chiral under $U(1)_Y\times U(1)$.
	
	We can repeat the same construction with $\{6,-4,-4,-3,-3,-3,-3,2,1,1,1,1,1,1,1,1,1,1,1,1\}$ of (\ref{best}). First, we append $\{4,-4\}$ to it and gather the three $-4$'s in one $U$ multiplet, the six $-3$'s in two $L$'s and the twelve $1$'s in two $Q$'s. Then, we  attribute the $6$ to $E$, $4$ to $\psi_1$ and $2$ to $\psi_3$. Again, for concreteness we can have
	
	\begin{align}
	[u_{ij}]=\hspace{2mm}\begin{blockarray}{ccccccc}
	N & \hspace{1mm} Q & \hspace{1mm} D & \hspace{2mm} L & \hspace{2mm} U & \hspace{1mm} E  \\
	\begin{block}{[cccccc]c}
	z_1 & 1 & 0 & -3 & 0 & 0 & \hspace{2mm} 1st\\
	-z_1 & 1 & 0 & -3 & -4 & 6 & \hspace{2mm} 2nd\\
	0 & 0 & 0 & 0 & 0 & 0& \hspace{2mm} 3rd\\ 
	\end{block}\\
	\end{blockarray},
	\end{align}

	\noindent and
	
		\begingroup
\setlength{\tabcolsep}{10pt} 
\renewcommand{\arraystretch}{1} 
    \begin{table}[ht]
	\begin{tabular}{ccc}
	\textrm{Fermions fields} & $U(1)_Y$ & $U(1)$ \\ \hline
	$\psi_1$ & 4 & 4 \\
	$\psi_2$ & -4 & 0\\
	$\psi_3$ & 2 & 2\\
	$\psi_4$ & -2 & 0\\ \hline
	\end{tabular}
	\end{table}
	\endgroup
	
	\noindent Note that we added opposite charges $z_1,-z_1\in\mathbb{Z}$ to the pair of right-handed neutrinos which does not spoil anomaly cancellations. As for the other extension (\ref{uextension}), the anomaly equations are satisfied for this one, in particular

    \begin{align}
    & \sum_{i,j}g_j\delta^{su(2)}_ju_{ij}=-2\times 2\times3+2\times 1\times 6=0,\\
    & \sum_{i,j}g_j\delta^{su(3)}_ju_{ij}=-3\times 4-2\times3+6\times 1=0.
    \end{align}
	
	These two composite anomaly-free extensions are particular cases of the extensions with the charges of the SM fermions under the additional $U(1)$ given by
	
	\begin{align}
	[u_{ij}]=\sigma(k_1[u_{ij}]+k_2[v_{ij}]+k_3[\delta_{i3}y_j]),
	\label{ucharges}
	\end{align}
	
	\noindent with $z_1,k_1,k_2,k_3\in\mathbb{Z}$, $\sigma\in S_3^6$ and $\delta_{i3}$ the usual Kronecker delta. We also need to add a chiral set with four Weyl fermions that have $U(1)_Y\times U(1)$ charges given by
	
	\begingroup
\setlength{\tabcolsep}{10pt} 
\renewcommand{\arraystretch}{1} 
    \begin{table}[ht]
	\begin{tabular}{ccc}
	\textrm{Fermions fields} & $U(1)_Y$ & $U(1)$ \\ \hline
	$\psi_1$ & 4 & 4$k_1$ \\
	$\psi_2$ & -4 & -4$k_2$\\
	$\psi_3$ & 2 & 2$k_1$\\
	$\psi_4$ & -2 & -2$k_2$\\ \hline
	\end{tabular}
	\end{table}
	\endgroup

	\noindent Note that these extensions are in a sense the simplest possible with additional charged chiral fermions because they are made of multiples of hypercharge which turns the cross anomaly equations into the cubic that are satisfied by construction. This implies that we can build $U(1)^m$ extensions from them as we did in (\ref{onefamilyU(1)^m}) and (\ref{U(1)^m}). A different approach would be to search other $U(1)$ extensions to the rows of (\ref{best}). The same construction can be done with the other partitions (\ref{1530})-(\ref{2025}), but for them one needs to add more Weyl fermions.

	\subsection{Conclusions}
	
	In this paper, we built composite anomaly-free $U(1)^m$ extensions of the SM. The availability of free parameters in (\ref{onefamilyU(1)^m}) and (\ref{U(1)^m}) that enter linearly in the formulas for the fermion charges and the large number of different classes of extensions may help other model builders interested in their use to solve problems of particle physics.
	
	\vspace{5mm}
	
	\noindent \textit{Acknowledgments}.  {\small   The author is supported by Funda\c{c}\~ao de Amparo \`a Pesquisa do Estado de S\~ao Paulo. He is grateful for the collaborations with Patrick Fox and Bogdan Dobrescu.
	}
	
	\newpage
	\providecommand{\href}[2]{#2}\begingroup\raggedright
\end{document}